\documentclass[manuscript]{aastex61}
\usepackage{txfonts}
\usepackage{latexsym,bm}
\usepackage{multirow}
\shorttitle{Modulation of GCRs over solar cycles}

\shortauthors{ Shen and Qin}

\begin{document}

\title{Modulation of Galactic Cosmic Rays in the Inner Heliosphere over Solar 
Cycles}

\correspondingauthor{G. Qin}
\email{qingang@hit.edu.cn}

\author[0000-0002-4148-8044]{Z.-N. Shen}
\affiliation{State Key Laboratory of Space Weather, National
Space Science Center, Chinese Academy of Sciences,
Beijing 100190, China}
\affiliation{College of Earth Sciences, University of
Chinese Academy of Sciences, Beijing 100049, China}

\author[0000-0002-3437-3716]{G. Qin}
\affiliation{School of Science, Harbin Institute of Technology, Shenzhen, 
518055, China; qingang@hit.edu.cn}

\begin{abstract}
The 11-year and 22-year modulation of galactic cosmic rays (GCRs) in the 
inner heliosphere are studied using a numerical model developed by Qin and Shen 
in 2017. Based on the numerical solutions of Parker's transport equations, 
the model incorporates a modified Parker heliospheric magnetic field, a locally 
static time delayed heliosphere, and a time-dependent diffusion coefficients model 
in which an analytical expression of the variation of magnetic turbulence 
magnitude throughout the inner heliosphere is applied. Furthermore, during solar 
maximum, the solar magnetic polarity is determined randomly with the possibility 
of $A>0$ decided by the percentage of the north solar polar magnetic field being 
outward and the south solar polar magnetic field being inward. The computed results 
are compared with several GCR observations, e.g., IMP 8, SOHO/EPHIN, Ulysses, 
Voyager 1 \& 2, at various energies and show good agreement. It is shown that our 
model has successfully reproduced the 11-year and 22-year modulation cycles.
\end{abstract}


\section{Introduction}
Galactic cosmic rays (GCRs) are modulated by interplanetary magnetic 
field (IMF) while transporting inside the heliosphere, and show a $\sim11$-year 
cycle variation \citep[e.g.,][]{McDonald1998,ShenAQin2016}. 
Furthermore, one observes a $\sim22$-year cycle of the intensities of GCRs with 
peak-like or plateau-like temporal profiles in solar minimum during negative or 
positive solar magnetic polarity, respectively. Theoretical and numerical models 
have successfully illustrated that the variation of GCRs in the heliosphere is 
caused by the modulation processes, including convection, diffusion, drifts, and 
adiabatic energy changes 
\citep[see, e.g., ][]{Parker1965,Zhang1999,Pei2010,Strauss2012,Potgieter2013,Zhao2014,Qin2017}. 

The temporal profiles of GCRs also show some short-term variations, 
including the $\sim27-$day solar rotation variations and irregular variations 
such as Forbush decreases \citep{Cane2000,Richardson2004,Alania2011,ShenAQin2016}. 
The $\sim27-$day variations of GCRs are caused by the passage of corotating 
interaction regions \citep[CIRs, ][]{Richardson2004}. The GCR depressions onset 
is thought to be related to the stream interfaces and the leading edges of CIRs 
\citep{Richardson1996}, and diffusion has a more important role than drifts in 
$\sim27-$day modulation of GCRs \citep{Guo2016}. In addition, the $\sim27-$day 
variation amplitudes of GCR during $A>0$ (magnetic field in the northern hemisphere 
directed outward the Sun) solar minima are larger than that during $A<0$ (magnetic 
field in the northern hemisphere directed toward the Sun) solar minima 
\citep{Richardson1999,Alania2001,Kota2001}. \citet{Gil2017} summarized that for 
the $\sim27-$day modulation of GCRs, the heliospheric current sheet (HCS) plays 
an important role during $A<0$ solar minima, and the latitude dependent solar 
wind velocity is more important during $A>0$ solar minima. Forbush decreases (FDs) 
are short-term decreases in GCR intensity followed by a substantially slower 
recovery \citep{Lockwood1971,Cane2000}. They are typically caused by the 
interplanetary counterparts of coronal mass ejections (CMEs) from the Sun. 
\citet{Cane2000} depicted a classical two-step FD, where the first step of the 
decrease is caused by the turbulent field generated behind the CME associated 
interplanetary shock and the second step is caused by the closed field line 
geometry of CME ejecta. \citet{Wawrzynczak2010} studied the temporal 
variabilities of the rigidity spectrum exponent of FD, and showed that the 
hardening rigidity spectrum was caused by the increased magnetic turbulence. 
Utilizing the measurements from the worldwide neutron monitor stations, 
\citet{Zhao2016} studied two prominent FD events during solar cycle 24 with comparable 
decrease magnitudes but distinctly different evolution of GCR energy spectrum 
and the energy dependence of recovery time. The different 
solar wind structures were thought to account for the difference \citep{Zhao2016}. 
While propagating outward, the corotating streams or CME generated transient 
solar wind flows usually merge 
to form various types of large interaction regions, e.g., the merged interaction 
regions \citep[MIRs, ][]{Burlaga1993,Cane2000} with the largest ones known as 
global MIRs (GMIRs). The GMIRs, which are considered to be related to CMEs 
\citep{Cane2000}, can cause large decrease in GCR intensity and are treated as 
propagating diffusion barriers in some modeling works 
\citep[e.g., ][]{Potgieter1992,LeRoux1995}. During the 11-year solar cycle, 
barrier modulation is dominant at solar maximum, and drift effects are thought 
to control modulation during solar minimum \citep{Cliver2013,Potgieter2013}. 
Note that CIRs do not contribute to the 11-year modulation cycle 
\citep{Potgieter2013}, while a series of GMIRs may be responsible for the 
long-term GCR modulation during the increasing phase of solar activity 
\citep{Potgieter1992,LeRoux1995}. 

Modulation in steady-state has been well studied by previous works 
\citep{Potgieter2013,Potgieter2014,Zhao2014}. \citet{Potgieter2014} studied the 
modulation of proton spectra with the PAMELA data from July 2006 to 2009, and 
they concluded that the recent solar minimum was ``diffusion dominated". In the 
work of \citet{Potgieter2014},
parameters in diffusion coefficients and drifts coefficients were adapted 
to observations \citep[see also][]{Potgieter2015,Raath2016}. \citet{Zhao2014} 
studied modulation of GCRs energy spectra during the past three solar minima
using an empirical diffusion coefficient model according to \citet{Zhang1999}. 
They found that decreased perpendicular diffusion in polar direction, 
which is in contrast to the assumption of enhanced diffusion in polar regions 
that was used to explain the observed Ulysses CR gradients 
\citep[see, e.g.,][]{Potgieter2000}, and increased parallel diffusion 
might be the reason for the record high-level of GCR intensity measured at Earth. 
Since the diffusion coefficients describe the scattering of GCRs by random 
fluctuations in the IMF, turbulence quantities are needed in the diffusion theory. 
In solar wind, the evaluation of turbulence are well described by 
magnetohydrodynamic (MHD) theory \citep{Marsch1989,Zhou1990}, and the turbulence 
transport throughout the heliosphere has been studied over the years 
\citep[e.g.,][] {Zank1993,Zank1996,Zank2012,Zank2017,Matthaeus1999,Smith2001,
Breech2008,Hunana2010,Oughton2011,Wiengarten2016}. With the turbulence transport 
models, the diffusion tensor is able to be calculated, 
\citep[e.g.,][]{Zank1998,Pei2010a,Engelbrecht2013a,Zhao2017}. Theoretical and 
numerical works have shown that drift coefficients can be reduced in the presence 
of turbulence,  \citep[e.g.,][]{Jokipii1993,Fisk1995,Giacalone1999,Candia2004,
Stawicki2005,Minnie2007,Tautz2012}, see also the first-order approach in 
\citet{Engelbrecht2017}. The reduced drift coefficients, which are obtained from 
fitting the simulation results \citep{Burger2010,Tautz2012}, are also used together 
with the turbulence transport theory to study the modulation of GCRs 
\citep[e.g.,][]{Engelbrecht2013a}. Using the Nearly Incompressible (NI) 
MHD turbulence transport model developed by \citet{Zank2017}, \citet{Zhao2017} 
also showed the effect of both weak and moderately strong turbulence on drift coefficients. 
Since there exists close coupling between turbulence, solar wind, and energetic 
particles, some work combine large-scale solar wind flow with small-scale 
fluctuations in a self-consistent way 
\citep[see, e.g., ][]{Usmanov2011,Usmanov2014,Usmanov2016,Wiengarten2015,Shiota2017} 
to study the spatial variations of the diffusion coefficients 
\citep[see, e.g., ][]{Chhiber2017}. Furthermore, using a diffusion coefficient 
model according to \citet{Giacalone1999}, \citet{Guo2016} studied the modulation 
of GCRs by CIRs at 1 au. They combined the small-scale 
turbulence transport with the MHD background for the simulation of cosmic-ray 
transport to show short-term modulation effects.

In order to describe the cosmic-ray modulation over long periods, time-dependent 
modulation processes are needed. \citet{Perko1983} used a one-dimensional numerical 
model to study the time-dependent modulation of GCRs for the first time.  
\citet{LeRoux1990} established a time-dependent two-dimensional numerical 
model, which incorporated a time-dependent drift model and the wavy HCS varying 
with time and propagating outward at the solar wind speed. Observations show 
that large cosmic-ray transient decreases are caused by MIRs and GMIRs 
\citep{Burlaga1993,Webber1993}. Therefore, some time-dependent modulation models 
used the combination of drifts and GMIRs \citep[e.g.,][]{Potgieter1993,LeRoux1995} 
to provide a complete 22-year modulation cycle. \citet{Ferreira2004} used a 
two-dimensional numerical model, and suggested a compound approach, 
which combined the effects of the changes in the HMF magnitude with drifts, 
also time-dependent current-sheet tilt angles. Thus, they established a time 
dependence in the diffusion and drift coefficients. This approach adopted the 
concept of propagating diffusion barriers \citep[see, e.g., ][]{Burlaga1993,LeRoux1995}, 
and gave lower values of diffusion coefficients and drifts for solar maximum than 
for solar minimum. Using the model established by \citet{Ferreira2004}, 
\citet{Ndiitwani2005} studied modulation of GCRs along the trajectory of Ulysses 
and concluded that drifts needed an additional decrease during solar maximum. 
Based on their works, \citet{Manuel2011} and \citet{Manuel2014} considered 
the dependence of diffusion coefficients on turbulence nature 
\citep[e.g., ][]{Teufel2002,Teufel2003,Shalchi2004,Minnie2007} and developed an 
improved compound approach. Such compound approaches were incorporated in numerical 
modulation model, and the computed results agreed well with observations at various 
energies. \citet{Bobik2012} established a 2D heliospheric modulation (HelMod) model 
with a re-scaled heliosphere which was divided into radially equally-spaced slices. 
In their work, the time-dependent modulation parameter $K_0$ in diffusion coefficients 
was determined by the modulation strength which was given by the force-field model 
\citep[FFM, see, e.g., ][]{Gleeson1968,Gleeson1971}. Then they fitted $K_0$ with 
the sunspot numbers and the neutron monitor counting rates for periods of low and 
high solar activities, respectively. \citet{Boschini2017} used this model 
to study modulation of GCRs with energy approximately larger than 0.5 GeV/nucleon, 
and the numerical results were consistent with the observations of PAMELA, AMS-02 
and Ulysses. 
\citet{Qin2017} used a new diffusion coefficients model NLGCE-F from \citet{Qin2014} 
to study the modulation of GCRs from 2006 to 2009. 
The NLGCE-F model was obtained by fitting the numerical solutions of 
nonlinear 
perpendicular \citep[NLGC,][]{Matthaeus2003} and parallel \citep[NLPA,][]{Qin2007} 
diffusion theories with polynomials, so it can help us calculate the diffusion 
coefficients directly without the iteration solution of integration equations set. 
The NLGCE-F model needs turbulence quantity throughout the heliosphere, which 
has been studied by turbulence transport models  
\citep[see, e.g.,][]{Zank1996,Zank2012,Zank2017,Breech2008,Pei2010a,Oughton2011,
Engelbrecht2013a}. 
However, \citet{Qin2017} only considered the variability of magnetic turbulence 
magnitude in the heliosphere for simplicity. They established an 
analytical expression, denoted as TRST, to describe the spatial variability of 
turbulence magnitude, which is consistent with the Ulysses, Voyager 1 \& 2 
observations. The turbulence model TRST was then inserted into the diffusion 
model NLGCE-F to establish a time and space dependence in the diffusion 
coefficients. 

The main purpose of this paper is to reproduce the $\sim11-$ and 
$\sim22-$year
solar cycle variation of GCRs using the numerical modulation model based 
on \citet{Qin2017}. In section 2, we show interplanetary conditions observed near 
Earth and briefly discuss the HMF polarity in solar maximum. In section 3, we 
discuss the GCR modulation model, including particle drifts and diffusion 
coefficients. 
Computation results are compared with observations of IMP 8, Ulysses, 
Voyager 1 \& 2 at various energies in section 4. Conclusions and discussion are 
presented in section 5.

\section{Interplanetary Conditions in the Inner Heliosphere}

\begin{figure}
\epsscale{1.} \plotone{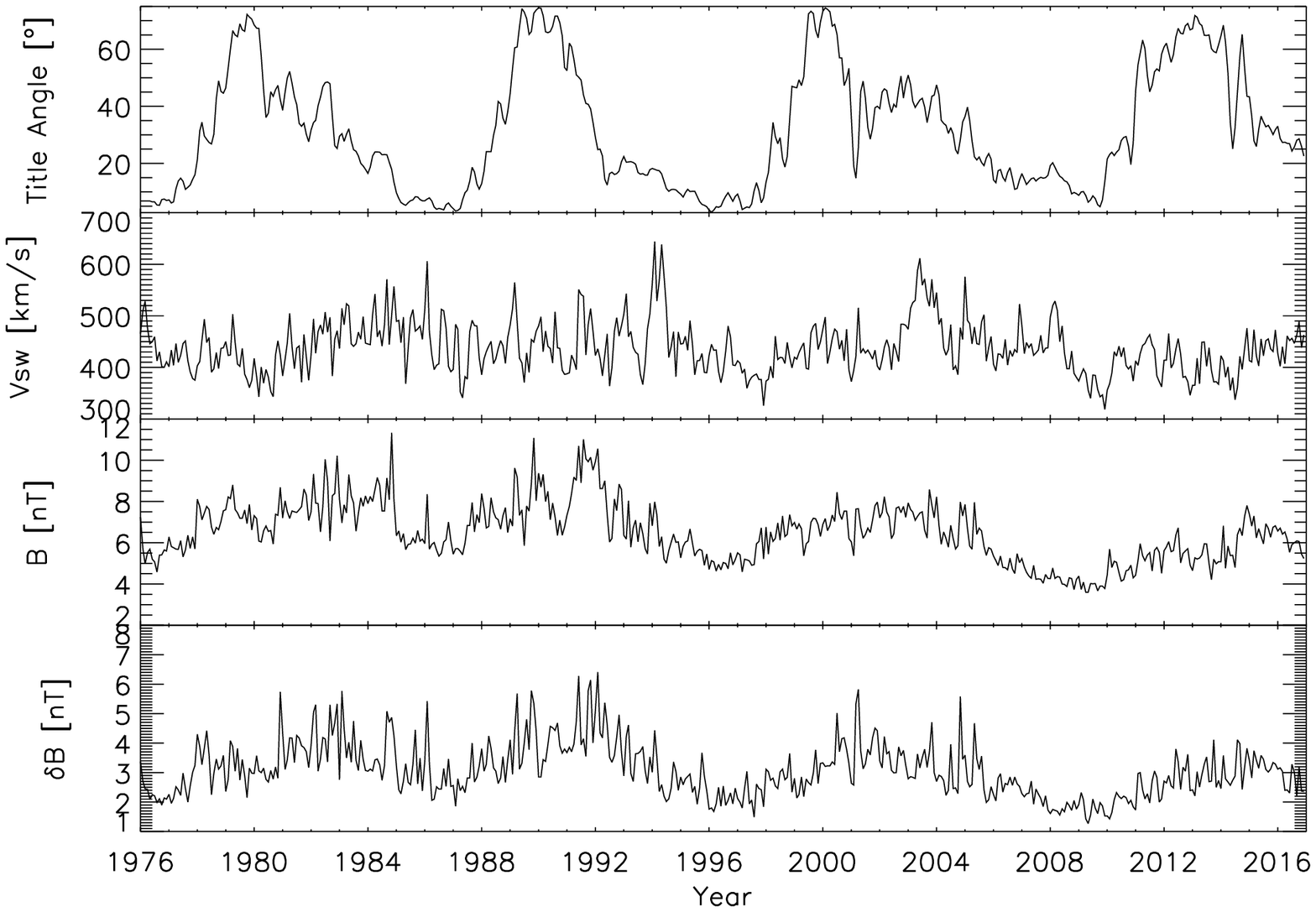}
   \caption{Interplanetary parameters at 1 au. Top panel shows the tilt 
angle of heliospheric current sheet from the WSO website (\url{wso.stanford.edu}) 
with the ``new" model. Second and third panels represent averaged solar wind velocity 
and averaged magnetic field strength for each month, respectively. 
Black line in the bottom panel means the magnetic turbulence magnitude calculated 
using method described in \citet{Qin2017}, and minutely HMF data from OMNI website 
(\url{omniweb.gsfc.nasa.gov}) is used to get more accurate result. 
}
   \label{fig:input}
 \end{figure}

Solar activities show 11-year cycles, heliospheric parameters embedded in the 
solar wind plasma also show cyclic variabilities related to solar cycles. 
In our model, several time-varying heliospheric parameters are needed to study 
the long-term modulation of GCRs. Figure \ref{fig:input} illustrates 
observations of interplanetary conditions as a function of time for the period 
1976-2016. Top panel shows the computed tilt angle $\alpha$ for the new model from 
Wilcox Solar Observatory (\url{wso.stanford.edu}). HCS is nearly flat during 
solar minimum and the maximum values of tilt angle reach above $70^\circ$ around 
solar maximum. Second and third panels are monthly averaged solar wind speed and 
HMF strength near Earth using data from OMNI website (\url{omniweb.gsfc.nasa.gov}). 
The solar wind speed does not show a clear 11-year cycle, while the HMF 
strength depicts an 11-year cycle. It is noted that in 2009 the HMF reached the 
lowest value of about 3 nT since 1963. The last panel means the square 
root of magnetic field variances (magnetic turbulence magnitude) calculated using 
the method described in \citet{Qin2017}, and minutely HMF data from OMNI website is 
used to get more accurate result. Note that solar wind data in months of November 
1982, December 1982 and January 1983 are not available, we use the linear 
interpolation method to get the monthly values of solar wind velocity. 

\begin{figure}
\epsscale{1.} \plotone{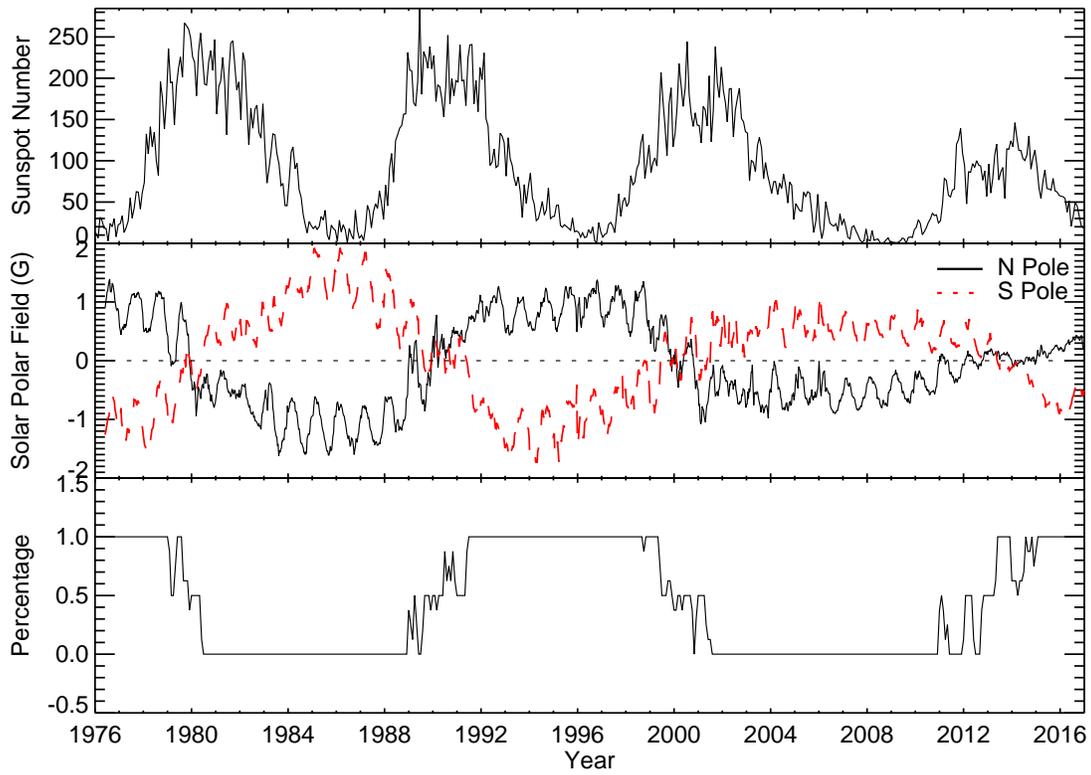}
   \caption{Top panel shows the monthly sunspot number from WDC-SILSO (\url{sidc.oma.be}). 
The second panel means the north (black line) and south (red dashed line) 
solar polar field from WSO (\url{wso.stanford.edu}). Bottom panel represents the 
possibility of $A>0$ for each month. See the text for more details. }
   \label{fig:field}
 \end{figure}

In addition, HMF changes its dipole polarity during solar maximum and resulting 
in a 22-year magnetic cycle. In Figure \ref{fig:field}, top panel means 
the sunspot number from the World Data Center SILSO (WDC-SILSO, \url{sidc.be/silso/}), 
the second panel represents the north (black solid line) and south (red dotted line) 
polar magnetic field strength, and the positive (negative) value of north solar 
polar field indicates HMF points outward (inward) in the northern hemisphere of 
the Sun, which is also noted as the $A>$ ($A<0$) polarity cycle. Hereafter we 
denote the north and south solar polar field strength as $B_N$ and $B_S$, respectively. 
The solar polar magnetic field reverses its polarity every $\sim11$ years 
and this causes the charge-sign dependent modulation and a 22-year solar modulation 
of GCRs. Therefore, the solar magnetic polarity plays an important role in the 
cosmic ray modulation theory. However, during solar maximum, it is difficult 
to identify the Sun's magnetic polarity, which is assumed as random. For any time 
interval of one month, we calculate the percentage $P_A$ of positive value in data 
set $B_N$ and $-B_S$, which is shown in the bottom panel of Figure \ref{fig:field}. 
We assume the possibility of $A>0$ in this month is $P_A$. From the bottom 
panel of Figure \ref{fig:field} we can see that $P_A$ varies in the range of [0, 1] 
around solar maximum, while $P_A=1$ or $P_A=0$ during the period away from solar 
maximum.
\section{Modulation Model and Numerical Methods}
Galactic cosmic rays are modulated by solar activities while transporting in 
the heliosphere, and the modulation processes are governed by the well known 
Parker transport equation (TPE) \citep{Parker1965},
\begin{equation}
\frac{\partial f}{\partial t} =-\left(\mathbf{V}_{sw}+ \left\langle{\mathbf{v}_d}\right\rangle
\right) \cdot\nabla f  +\nabla\cdot\left(\mathbf{K}_s\cdot\nabla f \right)+\frac{1}
{3}\left(\nabla\cdot\mathbf{V}_{sw}\right)\frac{\partial f}{\partial\ln p}, 
\label{eq:TPE}
\end{equation} 
where $f(\mathbf{r},p,t)$ is the omnidirectional cosmic-ray distribution function, 
with $\mathbf{r}$ the spatial position in a heliocentric spherical coordinate system, 
$p$ the particle momentum, $t$ the time. The cosmic-ray intensity $j= p^2f$. 
$\mathbf{V}_{sw}$ is the solar wind velocity, $\left\langle{\mathbf{v}_d}\right\rangle$ 
is the pitch-angle averaged drift velocity, and $\mathbf{K}_s$ is the symmetric 
part of diffusion tensor, which has three components in the field-aligned 
conditions, a parallel diffusion coefficient $\kappa_\parallel$ and two 
perpendicular diffusion coefficients, in the radial direction ($\kappa_{\perp r}$) 
and the polar direction ($\kappa_{\perp\theta}$). In this work, we assume the 
solar wind velocity is radial and the magnitude remains constant in the inner 
heliosphere. Thus $\left(\nabla\cdot\mathbf{V}_{sw}\right)>0$ and the last term 
on the right hand side represents adiabatic energy losses. 

The background HMF, which is embedded in the solar wind plasma, is usually assumed 
to have an Archimedean spiral due to the solar rotation \citep{Parker1958}. 
However, the standard Parker HMF is always modified over the polar regions of the 
heliosphere to scale down the excessive drift effects 
\citep{PotgieterEA89,Potgieter2013,Raath2016,Qin2017}. 
HMF observations at high latitude also show significant deviations from 
what the standard Parker model predicts \citep{Forsyth1996}. Thus several 
modifications of Parker HMF has been proposed, \citet{Jokipii1989} introduced
a small latitudinal component, \citet{Smith1991} added an additional azimuthal term, 
\citet{Fisk1996} included the footprint motion of HMF field lines to account for 
observations of recurrent energetic particle events and smaller 
cosmic ray intensities at higher latitudes \citep[see, e.g., ][]{Simpson1996, 
Zhang1997}. 
However, the Fisk type HMF is difficult to handle to some extent, and its existence 
is also debated \citep[e.g., ][]{Zurbuchen1997,Roberts2007,Sternal2011}, 
see the review in \citet{Potgieter2017}. Following \citet{Qin2017}, we use the 
modification proposed by \citet{Jokipii1989} in this work, and the modified 
expression can be written as \citep{Langner2004} 
\begin{equation}
\mathbf{B}=\frac{AB_0}{r^2}\left(\mathbf{e}_r + 
\frac{r\delta(\theta)}{r_s}\mathbf{e}_{\theta} -\frac{(r-r_s)\Omega\sin\theta}
{V_{sw}}\mathbf{e}_\phi\right)[1-2H(\theta-\theta')],
\label{eq:parker1}
\end{equation}
where $\mathbf{e}_r$, $\mathbf{e}_\theta$, and $\mathbf{e}_\phi$ are unit vectors 
in the radial, latitudinal, and azimuthal directions, respectively; $B_0$ is a 
constant that allows $\left|\mathbf{B}\right|$ equal to the HMF magnitude value 
at Earth; $A=1$ ($A=-1$) describes the $A>0$ ($A<0$) solar cycle and the HMF 
points outward (inward) in the northern hemisphere of Sun; $\delta(\theta)$ is 
the perturbation parameter, $r_s$ is the radius of the source surface where the 
HMF is assumed to point radially outward, $\Omega=2.66\times10^{-6}$ rad s${}^{-1}$ 
is the angular speed of Sun's rotation, $H$ is the Heaviside function, 
$\theta^\prime$ is the HCS latitudinal extent whose expression is given by 
\citet{Kota1983}, 
\begin{equation}
\theta^\prime = \frac{\pi}{2}-\arctan\left[\tan\alpha\sin\left(\phi+
\frac{\left(r-r_s\right)\Omega}{V_{sw}}\right)\right]. 
\label{hcs}
\end{equation}
According to \citet{Langner2004}, we can express the perturbation parameter as 
\begin{equation}
\delta(\theta)=\frac{\delta {}_m}{\sin \theta}
\end{equation}
to obtain $\nabla\cdot\mathbf{B}=0$. In order to avoid singularity, we use a 
reflective boundary condition near the poles, $\theta=2\theta_0-\theta$, 
for $\theta<\theta_0$ if $\theta<90^\circ$ or $\theta>\theta_0$ if 
$\theta>90^\circ$. In the work of \citet{Jokipii1989}, the value 
of $\delta(\theta)$ was suggested in the range of $10^{-3}-3\times10^{-3}$. 
\citet{Bobik2013} also studied the value of $\delta_m$, and $2\times10^{-5}$ 
was selected to compare with the observations in cycle 23. Thus $\delta_m$ may be 
different in each solar cycle. However, the variation of $\delta_m$ within the range 
suggested by \citet{Bobik2013} does not have much modulation effect in our model
\citep[see also,][]{Boschini2017}. Therefore, in this study we set 
$\delta {}_m=2\times 10^{-5}$ 
\citep{Bobik2013,Boschini2017}, $\theta_0= 2.5^\circ$ if $\theta<90^\circ$ and 
$\theta_0= 177.5^\circ$ if $\theta>90^\circ$. It is shown that this modification 
can significantly enhance the magnetic field intensity and reduce the drift velocity 
at large radial distance in polar regions of the heliosphere without influence 
the equatorial regions obviously \citep{Bobik2012,Bobik2013,Qin2017}. 

\begin{figure}
\epsscale{1.} \plotone{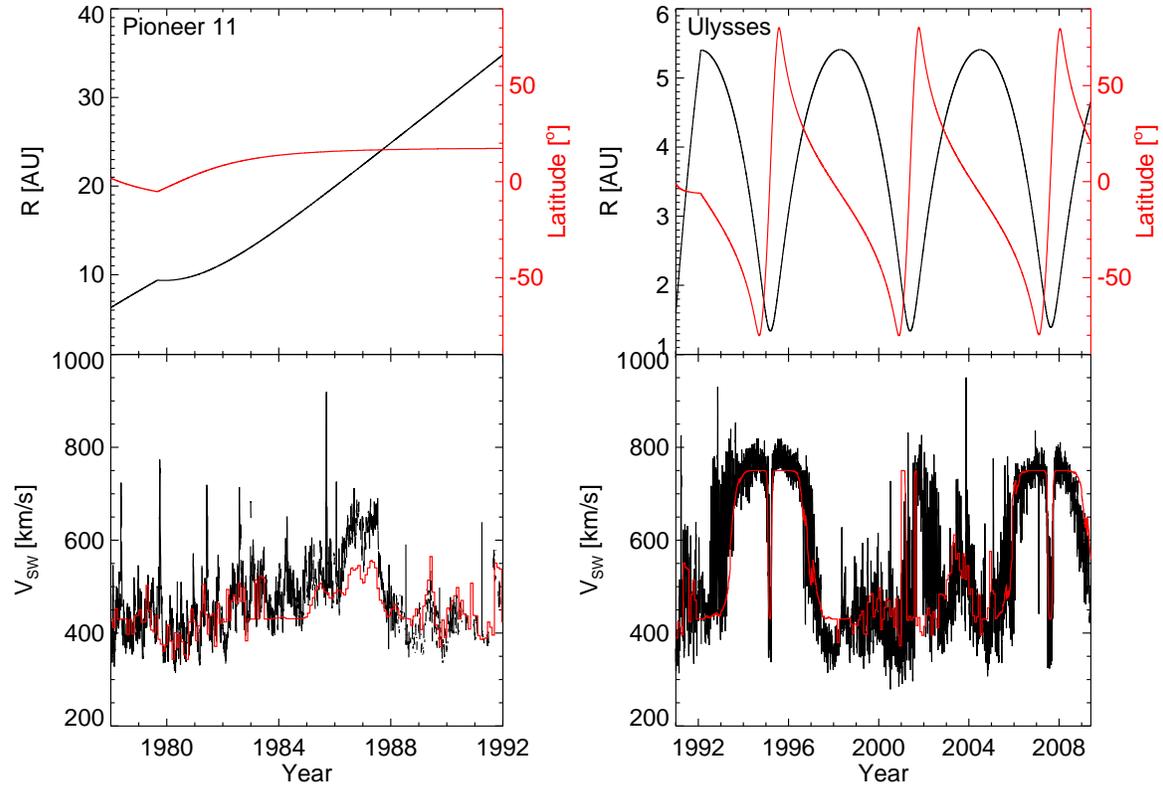}
   \caption{Left: Top panel shows the radial distance (black line) and heliographic 
latitude (red line) of Pioneer 11. Bottom panel means the comparison between the 
results of Equation (\ref{eq:SWmodel}) (red line) along the trajectory of 
Pioneer 11 and the daily solar wind speed observed by Pioneer 11 (black line). 
Right: The same to Figure \ref{fig:solarwind} (left) except that for Ulysses.
}
   \label{fig:solarwind}
 \end{figure}

According to \citet{Sheeley1997}, the solar wind speed accelerates from zero to 
a constant within 0.3 au from the Sun. During solar minimum, observations have 
shown that solar wind speed increases from $\sim 400$ km s${}^{-1}$ in the 
equatorial plane to $\sim 800$ km s${}^{-1}$ in high latitudes 
\citep{McComas2002, Zurbuchen2007}. However, such simple pattern does not exist 
anymore during solar maximum \citep{Heber2006}. Some works use a hyperbolic 
function to present such variability during solar minimum, and the function can be 
expressed as \citep[see, e.g., ][]{Hattingh1998,Heber2006,Potgieter2013} 
\begin{equation}
\mathbf{V}_{sw}(r,\theta) =V_{model}\mathbf{e}_r = V_0\left\{1-
	\exp\left[\frac{40}{3}\left(\frac{r_s-r}{r_0}\right)\right]
\right\}\left\{1.475\mp 0.4\tanh\left[6.8(\theta-\frac{\pi}{2}\pm\xi)\right]\right\}
	\mathbf{e}_r, \\
\label{eq:solarwind}
\end{equation}
with $V_0=400 \textrm{km/s}$, $r_0=1$ au and $\xi=\alpha+15\pi/180$. 
The top and bottom sign correspond to the northern and southern hemisphere, 
respectively. Note that coefficients in Equation (\ref{eq:solarwind}) are 
obtained by comparing with the observations of Ulysses \citep[see, e.g., ][]{Hattingh1998}. 
\citet{Potgieter2013a} classified the solar activity in terms of $\alpha$, with 
$\alpha\leq 30^\circ$, $30^\circ<\alpha\leq 60^\circ$ and $60^\circ<\alpha\leq 90^\circ$ 
represents periods of low, moderate and high solar activities respectively. 
In this work, we assume the solar wind speed has no latitudinal variation during 
periods of moderate and high solar activities with value $V_{1au}$ extracted from 
OMNI data set \citep[see, e.g., ][]{Bobik2012}. Thus the solar wind speed model 
used in this work can be expressed as follows, 
\begin{equation}
V_{sw}= 
\left\{
\begin{array}{ll}
V_{model}, \text{for $\alpha\leq 30^\circ$} \\
V_{1au}, \text{for $\alpha > 30^\circ$}. \\
\end{array}
\right.
\label{eq:SWmodel}
\end{equation}
Figure \ref{fig:solarwind} (left) shows the result of Equation (\ref{eq:SWmodel}) 
along the trajectory of Pioneer 11. The radial distance (black line) and 
heliographic latitude (red line) of Pioneer 11 are illustrated in the top panel. 
Black and red lines in the bottom panel indicate daily solar wind speed observed by 
Pioneer 11 and monthly solar wind speed predicted by Equation (\ref{eq:SWmodel}),
respectively,
along the trajectory of Pioneer 11. It is shown that Equation (\ref{eq:SWmodel})
can show the trend of solar wind speed in the latitude dependence comparying with
Pioneer 11 observations during solar minimum. 
We also show the daily solar wind speed observations of Ulysses (black line) 
and the result of Equation (\ref{eq:SWmodel}) along the trajectory of Ulysses 
(red line) in the bottom panel of Figure \ref{fig:solarwind} (right). The 
characteristics of solar wind velocity during solar maximum and minimum can be 
seen clearly. Better agreement between the Ulysses observation and Equation
(\ref{eq:SWmodel}) is obtained because the coefficients in Equation 
(\ref{eq:SWmodel}) are
got by comparying with Ulysses observations \citep[e.g.,][]{Hattingh1998}.
Note that for simplicity, while solving the TPE Equation (\ref{eq:TPE}) 
numerically in each step we assume the magnitude of solar wind as a constant with 
the value calculated with Equation (\ref{eq:SWmodel}). 

The drift coefficient can be suppressed by magnetic turbulence in solar wind plasma
 \citep[see, e.g., ][]{Jokipii1993,Fisk1995,Giacalone1999,Candia2004,Stawicki2005,
Minnie2007,Tautz2012}. 
However, it is complicated to use the reduction of drift effects self-consistently 
\citep[see, e.g.,][]{Bieber1997, Burger2010, Tautz2012} or in Ad hoc form 
\citep[see, e.g.,][]{Burger2000,Potgieter2013,Vos2016,Nndanganeni2016} in modulation 
works. The effects of turbulence on GCR drifts are far from complete 
to be understood. Therefore, we use the weak scattering drift coefficient for 
simplicity in this work. 
The general weak-scattering drift velocity can be written as \citep{Jokipii1977}
\begin{equation}
\langle \mathbf{v}_d \rangle=q\frac{P\beta}{3}\nabla\times\left(\frac{\mathbf{B}}
	{B^2}\right),
\end{equation}
with q the particle charge sign, $P$ the rigidity of particle, $\beta$ the ratio 
between the speed of particle $v$ and that of light, and $B$ the magnitude of 
modified Parker HMF. The drift velocity is usually divided into two 
components \citep{Burger1989}, i.e., the combination of gradient and curvature 
drifts $\mathbf{v}_{gc}$ which are caused by the large scale HMF, and the current 
sheet drift $\mathbf{v}_{ns}$ when particles across or transport near the HCS. 
Gradient and curvature drifts could lead to a $\sim22-$year cycle of GCR 
intensity and the charge-sign dependent modulation \citep{Jokipii1977}. Positively 
charged GCRs mainly drift inward from polar regions during $A>0$ polarity solar 
cycles, and mainly drift inward along the HCS in equatorial regions during $A<0$ 
polarity solar cycles. Thus drifts will produce flat and sharp temporal profiles 
of positively charged GCR intensity during the $A>0$ and $A<0$ solar minima, 
respectively. This effect reverses for negatively charged GCRs. 
The detailed description and derivation of drifts for the modified Parker 
HMF Equation (\ref{eq:parker1}) can be found in \citet{Qin2017}. 

Magnetic turbulence in solar wind plasma causes diffusion of cosmic rays 
parallel and perpendicular to the background HMF. Scattering theories have been 
developed to describe the properties of the diffusion coefficients \citep[see 
e.g.,][]{Jokipii1966,Matthaeus2003,Qin2007,Qin2014}. Following \citet{Qin2017}, 
we use the NLGCE-F model which is in the polynomial form by fitting
the numerical results from simultaneously solving nonlinear theories of parallel
and perpendicular diffusion \citep{Qin2007, Qin2014}. 
The expressions of the NLGCE-F model are 
as follows, 
\begin{equation}
\ln\frac{\lambda _\sigma}{\lambda_{slab}} =\sum\limits_{i = 0}^{n_{\sigma 1}} 
a_i^\sigma\left(\ln\frac{R_L}{\lambda_{slab}}\right)^i
\label{eq:diffusion}
\end{equation}
with
\begin{eqnarray}
a_i^\sigma  &=& \sum\limits_{j=0}^{n_{\sigma 2}}b_{i,j}^\sigma\left(\ln\frac{E_{slab}}
{E_{total}}\right)^j\\
b_{i,j}^\sigma&=&\sum\limits_{k = 0}^{n_{\sigma 3}}c_{i,j,k}^\sigma \left(\ln
\frac{\delta B^2}{B^2} \right)^k\\
c_{i,j,k}^\sigma&=&\sum\limits_{l=0}^{n_{\sigma 4}}d_{i,j,k,l}^\sigma 
\left(\ln\frac{\lambda_{slab}}{\lambda_{2D}}\right)^l,
\end{eqnarray}
where $\sigma$ means $\perp$ or $\parallel$, diffusion coefficients 
$\kappa_\sigma=\frac{\lambda_\sigma v}{3}$ 
and we assume $\kappa_{\perp r}=\kappa_{\perp\theta}$ in this work, 
$\lambda_{slab}$ is the spectral bend-over scale of the slab component of 
turbulence and $\lambda_{2D}$ is that of the 2D component, 
$E_{slab}=\left\langle{\delta B^2}_{slab}\right\rangle$ means the magnetic 
turbulence energy in the slab component and 
$E_{total}=\left\langle{\delta B^2}\right\rangle$ is the total magnetic 
turbulence energy, $R_L$ means the particle's gyro-radius, and $\delta B/B$ 
represents the turbulence level. The polynomial order $n_{\sigma i}$ and 
coefficients $d_{i,j,k,l}^\sigma$ in Equation (\ref{eq:diffusion}) are obtained 
from \citet{Qin2014}, and the computer code for the NLGCE-F diffusion 
coefficients model can be found on the website at 
\url{http://www.qingang.org.cn/code/NLGCE-F}.
According to \citet{Qin2014}, input parameters should vary within specified 
ranges, shown as follows, to ensure the validity of NLGCE-F model,
\begin{eqnarray}
1&\lesssim&\frac{\lambda_{slab}}{\lambda_{2D}}\lesssim 10^3,\\
10^{-3}&\lesssim&\frac{E_{slab}}{E_{total}}\lesssim 0.85,\\
10^{-4}&\lesssim&\frac{\delta B^2}{B^2}\lesssim 10^2,\\
10^{-5}&\lesssim&\frac{R_L}{\lambda_{slab}}\lesssim 6.3.
\end{eqnarray}
Note that, to insure the values of input parameters in Equation 
(\ref{eq:diffusion}) vary within valid ranges, the particle's energy should be 
not much more than $10$ GeV and the modulation boundary of the numerical model 
should be set inside the termination shock (TS) \citep{Qin2017}. 

Characteristics of HMF and turbulence in solar wind plasma throughout the 
heliosphere are needed to calculate diffusion coefficients with the NLGCE-F model.
Considering two-component model of turbulence \citep{MatthaeusEA90} in solar wind, 
turbulence quantities have been studied by turbulence transport models (TTMs)
\citep{Zank1996,Zank2012,Zank2017,Breech2008,Pei2010a,Oughton2011,Engelbrecht2013a}. 
To some extent, it is complicated to apply the TTMs in long-term modulation of 
GCRs. In addition, the turbulence parameters in solar wind, such as 
$\lambda_{slab}$ and $\lambda_{2D}$, can not be observed by spacecraft directly 
and should be studied with some theoretical work, so their detailed knowledge is 
not complete \citep[e.g.,][]{MatthaeusEA90, Adhikari2017}. Therefore, we set 
them in simple forms according to some study for solar wind near Earth. 
Observational studies \citep{Osman2007,Dosch2013} have indicated that 
$\lambda_{slab}$ is about a factor of two larger than $\lambda_{2D}$. According 
to new results from multi-spacecraft measurements \citep{Weygand2009,Weygand2011}, 
for the slow solar wind, the correlation scale ratio of slab to 2D is around 2.6 
while in the fast wind the ratio is around 0.7. Observations found that the ratio 
of the 2D and slab turbulence energy is about 80\%:20\% \citep{Matthaeus1990,Bieber1994}, 
and the ratio can be different in slow and fast solar wind \citep{Dasso2005}. 
In solar wind and solar corona, the plasma beta ($\beta_p$) is of the order of 
$\beta_p\sim1$ and $\beta_p\ll 1$, and the 2D and slab energy ratio 80\%:20\% is 
usually used in some theoretical work based on this assumption 
\citep{Zank1992,Zank1993,Hunana2010}. In addition, the correlation scales and 
energies of 2D and slab turbulence may also vary with solar cycles, but it still 
needs some further study. Therefore, in this work, we 
set ${\lambda_{slab}}/{\lambda_{2D}}=2.6$, $\lambda_{slab}=0.02r$ \citep{Qin2017}, 
$E_{slab}/E_{total}=0.2$, and only consider the spatial and temporal changes of 
magnetic turbulence magnitude $\delta B$ in the heliosphere for simplicity. 

Note that,  NLGC has also been 
derived into many analytical expressions, e.g., \citet{Zank2004} and \citet{Shalchi2004}, 
which are used in many simulation models with the parallel diffusion coefficient from
the quasilinear theory 
\citep[QLT, ][]{Jokipii1966} as input. However, QLT is not very accurate 
compared to the simulation results because of the nonlinear effect 
\citep[e.g.,][]{Qin2002}. Based on the NLGC 
theory, \citet{Qin2007} combined the nonlinear theory of parallel and perpendicular 
diffusion together to obtain the NLGC-E model, which is shown to agree better with 
simulation results than original NLGC with parallel diffusion from QLT as input. 
Besides, NLGCE-F also solves the case that NLGC fails 
in pure slab or pure 2D turbulence. 

\begin{figure}
\epsscale{1.} \plotone{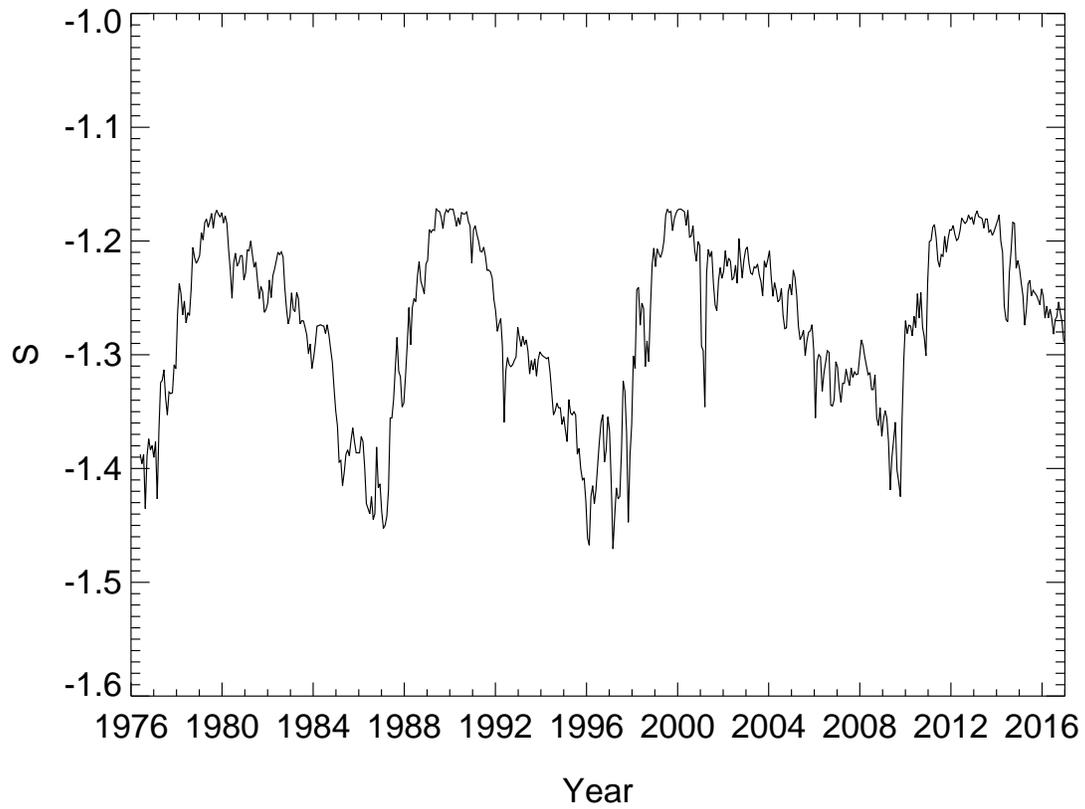}
   \caption{$S$ in Equation \ref{eq:S} varies with solar cycles.}
   \label{fig:spec}
 \end{figure}

For the radial dependence of $\delta B$, some studies use analytical expressions 
to approximate the solutions of TTMs 
\citep[see, e.g.,][]{Zank1996,Burger2008,Effenberger2012,Ngobeni2014,Strauss2017}, 
and the latitudinal dependence of $\delta B$ can be inferred from the magnetic field 
observations of Ulysses \citep{Perri2010}. Based on those studies, \citet{Qin2017} 
utilized an analytical expression, denoted as TRST, to describe the spatial 
variations of $\delta B$ as follows, 
\begin{equation}
\delta B = \delta B_{1 \mathrm{au}}R^S\left(\frac{1+\sin^2\theta}{2}\right),
\label{eq:turbulence}
\end{equation}
where $\delta B_{1 \mathrm{au}}$ means the magnetic turbulence magnitude near 
Earth, $R=r/r_0$ with $r_0=1$ au, and $S$ varies as a function of the HCS tilt 
angle $\alpha$ according to \citet{Qin2017}, 
\begin{equation}
S=-1.56+0.09\ln\frac{\alpha}{\alpha_c},
\label{eq:S}
\end{equation}
where $\alpha_c=1^\circ$. Figure \ref{fig:spec} shows $S$ varies 
with time. It is clearly shown that $S$ has an 11-year cycle. $\delta B$ decreases 
as $\sim r^{-1.19}$ and $\sim r^{-1.4}$ during solar maximum and solar minimum, 
respectively. We can infer from the prediction of WKB theory that $\delta B$ 
decays as $r^{-1.5}$. In the interplanetary space, stream shear, shock wave and 
pickup ions can be sources of turbulence \citep{Zank1996,Smith2001}. 
\citet{Zank1996} pointed out that the dissipation rate of turbulence energy is 
proportional to the source of turbulence. \citet{Adhikari2014} also verified that 
the temporally varying sources of turbulence depend on heliocentric distance, and 
all sources should be considered to fit the observations well. Considering all 
sources of turbulence, theoretical works \citep[see, e.g., ][]{Zank1996,Smith2001,Adhikari2014} 
have shown that the dissipation rate of turbulence energy is slower than that 
predicted by WKB theory. The parameters in TRST model are choosen to fit the 
spacecraft observations, and the complicated processes considered by TTMs have 
not been studied in this work. However, the dissipation rate of turbulence energy 
given by the TRST model is also slower than that predicted by WKB theory. It has 
been shown by \citet{Qin2017} that the TRST model agrees well with the magnetic 
turbulence magnitude observed by Ulysses, Voyager 1 and Voyager 2 measurements. 

\begin{figure}
\epsscale{1.} \plotone{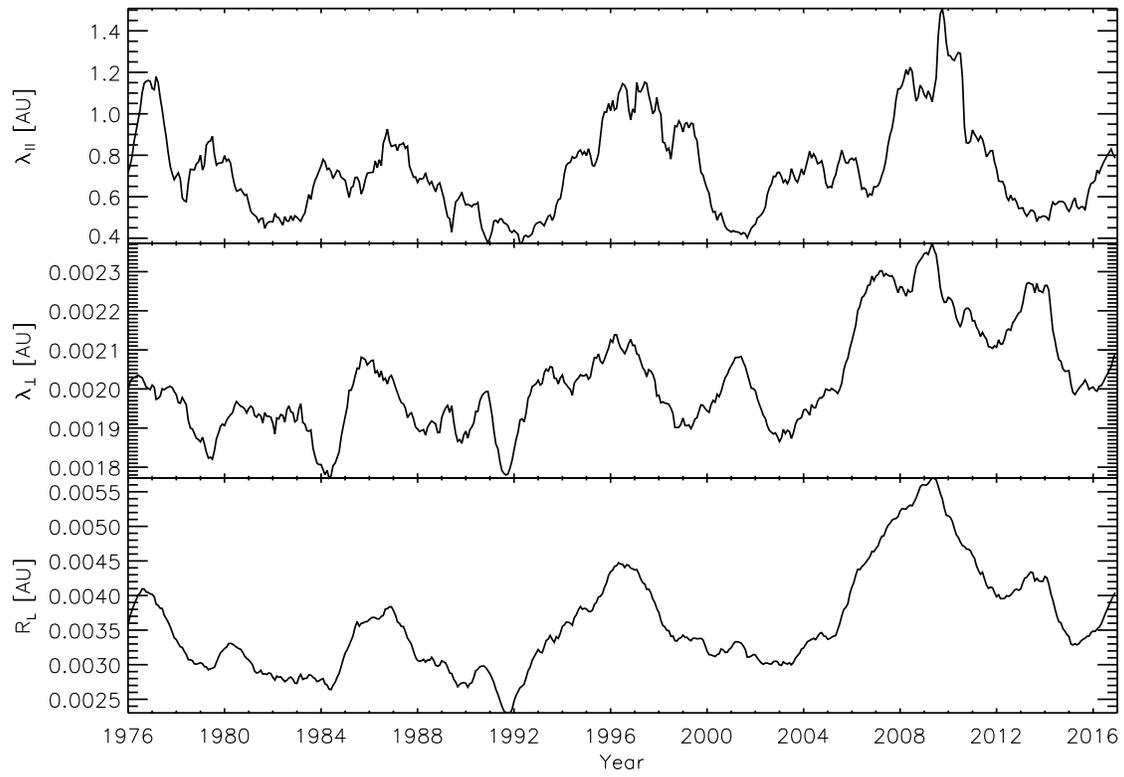}
   \caption{Computed monthly parallel and perpendicular mean-free paths 
as well as the particle's gyro-radius for 1 GV proton near Earth. The 13-month 
smoothed results are presented in this figure as the original values show too 
many spikes.}
   \label{fig:lambda}
 \end{figure}

We are able to compute diffusion coefficients at any time and location 
using diffusion coefficients model Equation (\ref{eq:diffusion}) with the modified 
Parker HMF Equation (\ref{eq:parker1}) and the TRST model. The input parameters, 
i.e., HMF strength at 1 au, $\delta B_{1 \mathrm{au}}$, and HCS tilt angle $\alpha$ 
can be obtained from observations near Earth. This diffusion coefficients model 
has been tested by reproducing the time-varying proton spectra observed by PAMELA 
during the past solar minimum \citep{Qin2017}. In this work, we would continue to 
test it with GCR flux observed by varies spacecraft, e.g., Ulysses, 
Voyager 1, \&2, over solar cycles. As an example, we show the computed monthly 
parallel and perpendicular mean-free paths as well as the particle's gyro-radius, 
which is equivalent to the particle's drift scale with the assumption of weak 
scattering, for 1 GV proton near Earth in the top, middle and bottom panel of 
Figure \ref{fig:lambda}, respectively. As the original results have shown too 
many spikes, we present the 13-month smoothed values in this figure. 
The $\sim 11$-year variation cycles are identifiable in the temporal 
profiles of $\lambda_{\parallel}$ and $R_L$, both of which reach quite  
low levels around solar maxima. The drift scale, as the turbulence effect not 
considered, is proportional to $B^{-1}$ and anti-correlated with solar 
activity. The profile of $\lambda_{\perp}$ presents some different 
variabilities, and there is no clear relationship between $\lambda_{\parallel}$ 
and $\lambda_{\perp}$. However, in the recent solar minimum, both  
$\lambda_{\parallel}$ and $\lambda_{\perp}$ show a large increase. 
The variations of turbulence magnitude $\delta B$ and IMF magnitude $B$ can 
cause the solar cycle variations of $\lambda_{\parallel}$ and 
$\lambda_{\perp}$. For a fixed $B$, the decreased $\delta B$ 
can result in an increase in $\lambda_{\parallel}$ and a decrease in 
$\lambda_{\perp}$. On the other hand, for a fixed $\delta B$, 
the decreased $B$ can result in a decrease in $\lambda_\parallel$ and an 
increase in $\lambda_\perp$. The  
relationship between parallel and perpendicular diffusion coefficients over solar 
cycles depends on the details of variations of $\delta B$ and $B$.
It is noted that \citet{Chhiber2017} also studied the influence of solar activity on
CR parallel and perpendicular diffusion with the modeling background magnetic field and
turbulence, they found that an increasing $\delta B$ leads to a decrease in 
$\lambda_{\parallel}$ and an increase in $\lambda_{\perp}$. We think it is
reasonable to have different relationship between parallel and perpendicular diffusion
coefficients using different $\delta B$ and $B$.

The real solar modulation boundary might be located beyond the heliopause 
\citep[e.g.,][]{Zhang2015}. Voyager 1 was assumed to have reached the very local 
interstellar medium in August 2012 \citep{Webber2013}, the very local interstellar 
spectrum (LIS) is able to be constructed using proton flux of Voyager 1 and that 
of PAMELA with energy larger than 30 GeV where the modulation effects can be 
neglected \citep{Potgieter2013,Vos2015,Bisschoff2016}. 
In addition, observations \citep{Stone2005,Stone2008} indicated that 
Voyager 1 and 2 reached the TS in 2004 and 2007 at about 94.0 AU and 83.7 AU, 
respectively. Thus the TS may have an asymmetric structure and a temporal position. 
The asymmetric structure can also be seen in some MHD models 
\citep[e.g., ][]{Opher2009,Florinski2009}, and has identifiable modulation 
effects on GCRs \citep{Langner2005}. The variations of the solar wind speed and 
density can result in a time-dependent TS position 
\citep[see, e.g., ][]{Whang1995,Whang2004,Wang1999,Webber2005,Webber2011,
Washimi2011,Richardson2011,Richardson2012}, and the TS position shows a latitude 
dependence \citep{Webber2011,Washimi2011}. However, the temporal profiles of the 
TS position shown by these theoretical works are somehow different. Thus the 
realistic TS position is complicated to be constructed considering its time and 
latitude dependence. As has been shown by the results of \citet{Manuel2014}, the 
TS and heliopause positions are important parameters in the modulation of GCRs. 
Considering a fixed heliopause position and a symmetric heliosphere, 
\citet{Manuel2015} 
studied the modulation effects of a time-varying sinusoidal TS position and a fixed one. 
Little difference was shown for the modulation of GCRs near Earth, but the 
time-dependent TS position (i.e., the time-dependent inner heliosheath thickness) 
had significant effects on the temporal variation of GCR intensity near TS. Since 
modulation effects in the outer heliosphere have not been included in our 
modulation model, following \citet{Qin2017} we set a symmetric outer boundary of 
modulation inside the termination shock at 85 au and use an input spectrum 
according to the observations of Voyager 2 at 85 au from \citet{Webber2008}. 
Following \citet{Zhang1999}, we express the GCR source at 85 au as follows, 
\begin{equation}
j_{s}=j_0 p_0^{2.6} p\left(m_0^2c^2+p^2\right)^{-1.8}
\label{eq:lis}
\end{equation}
where $j_0=1.17\times 10^4$
$\mathrm{m}^{-2}\mathrm{s}^{-1}\mathrm{sr}^{-1}\mathrm{(GeV/nuc)}^{-1}$ to fit 
Voyager 2 observations \citep{Qin2017}, 
$p_0=1~\mathrm{GeV}/c$ and $m_0$ means the rest mass of a proton.

In this work, we use the time-backward Markov stochastic process method 
which is proposed by \citet{Zhang1999} to solve the well known Parker transport 
equation. The Parker transport equation can be written in terms of a set of 
equivalent stochastic differential equations (SDEs) given by 
\citep{Zhang1999,Pei2010,Strauss2011,Kopp2012}
\begin{equation}
dx_i=A_i(x_i)ds+\sum_{j}B_{ij}(x_i) \cdot dW_j,
\end{equation} 
with $i\in{(r,\theta,\phi,p)}$, $x_i$ are the Ito processes \citep{Zhang1999}, 
$s$ is the backward time, and $\mathrm{d}W_i$ satisfy a Wiener process given by the 
standard normal distribution \citep{Pei2010,Strauss2011}. For a general HMF with 
a meridional component, the matrix components $B_{ij}$ are given by \citet{Pei2010} 
\citep[see also][]{Kopp2012}. For the components of vector $\mathbf{A}$ and the 
specific SDEs in spherical coordinate, please refer to \citet{Qin2017}. We trace a 
number of 
pseudo-particles from the observation location outward to the outer boundary and 
get particle intensities with the GCR source spectrum. We assume a locally static 
heliosphere for each month, and interplanetary conditions 
(e.g., $V_{sw}$, $B$, $\delta B$, $\alpha$, $P_A$) 
in position $\mathbf{r}$ at time $t$ are decided by the states at the source 
surface $r_s$ at an earlier time \citep{Potgieter2014}, thus we use a time-delayed 
heliosphere according to \citet{Qin2017}. We solve the TPE at monthly intervals. 

\section{Modeling Results}

While studying the modulation of GCR over solar cycles, we still need to know the 
characteristics of energy spectra both in solar minimum and solar maximum. 
In our previous work \citep{Qin2017}, we have successfully reproduced the proton 
spectra observed by PAMELA measurements as a function of time in the past $A<0$ 
solar minimum. However, modeling work during solar maximum remains a challenge. 
During the maximum phase of each cycle, the heliosphere is more complex due to 
the high-level solar activities, and the Sun's polar magnetic field changes 
sign (i.e., the sign of $A$) frequently. For simplification, we assume the solar 
wind speed has no latitudinal component during solar maximum according to the 
observations of Ulysses \citep{Heber2006} and use a time delayed heliosphere 
\citep{Qin2017}. In addition, for each pseudo-particle in each time step, the 
solar magnetic field polarity $A$ is decided as $A=1$ or $A=-1$ randomly 
according to the percentage $P_A$. 

\begin{figure}
\epsscale{1.} \plotone{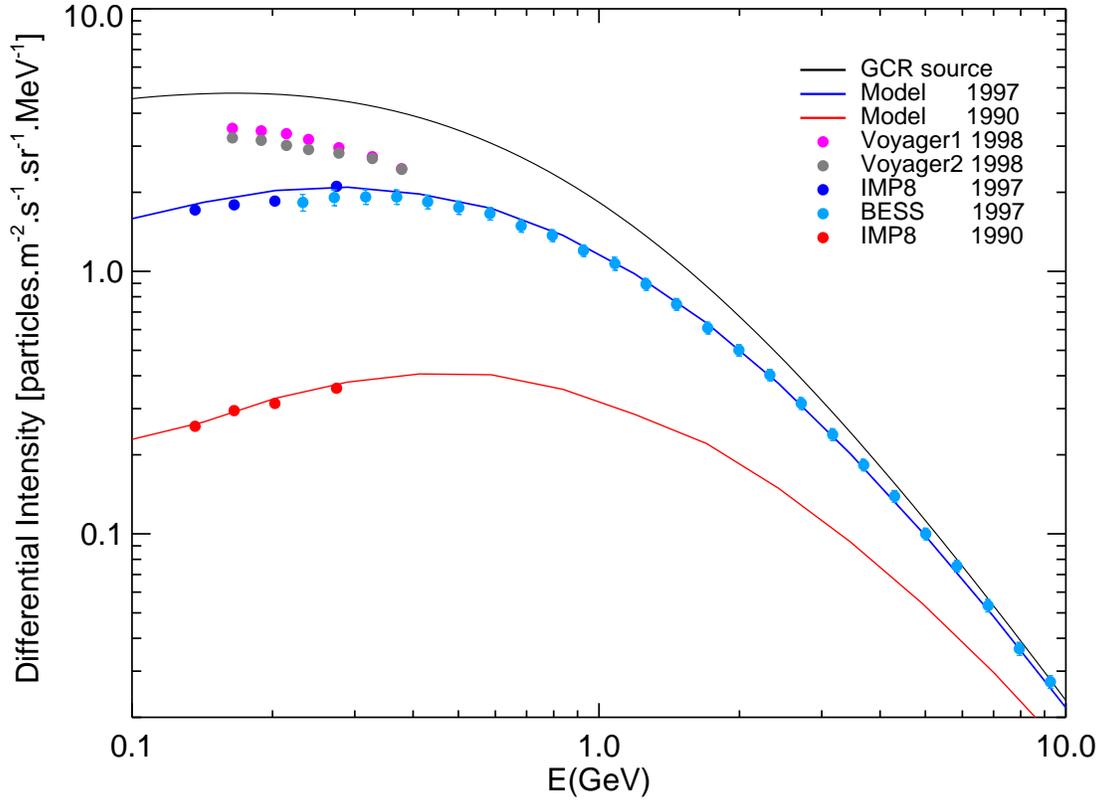}
   \caption{Computed GCR proton energy spectra at Earth near solar minimum in 1997 
	(blue line) and near solar maximum in 1990 (red line), as well as observations of 
	various instruments. The black line means GCR source used in this work 
	\citep{Qin2017}. Magenta and gray solid circles denote Voyager 1 and 
	Voyager 2 observations in 1998 reported by \citet{Webber2006}. Blue and sky 
	blue solid circles represent IMP 8 data in 1997 from \citet{Webber2003} and 
	data from the BESS magnetic spectrometer in 1997 \citep{Shikaze2007}. Red 
	solid circles refer to yearly averaged GCR proton flux, which has 
	been processed to get the GCR background using the despiking algorithm 
	\citep{Qin2012}, from IMP 8 in 1990.}
   \label{fig:imp8}
 \end{figure}

Figure \ref{fig:imp8} shows an example of the modulated proton energy spectra 
near solar maximum in 1990 (red line) and near solar minimum in 1997 (blue line). 
Black line represents the GCR source used in this work \citep{Qin2017}. Blue and 
sky blue solid circles mean IMP 8 observations in 1997 from \citet{Webber2003} 
and BESS magnetic spectrometer observations in 1997 from \citet{Shikaze2007}, 
respectively. Red solid circles refer to yearly averaged GCR proton flux observed 
by IMP 8 in 1990. Note that the original 30-minute resolution proton flux data 
(from CDAWeb: \url{cdaweb.sci.gsfc.nasa.gov}) has been processed to get the GCR 
background using the despiking algorithm \citep{Qin2012}. The computed energy 
spectrum in solar maximum is harder than that in solar minimum, and the 
results of our model are consistent with these GCR observations. In this figure, 
the spectra observed by Voyager 1 (magenta circles) and Voyager 2 (gray circles) 
in 1998 reported by \citet{Webber2006} are shown even softer than the GCR source, 
which can be hardly reproduced by our model. Therefore, in the following, instead 
to study the energy spectra, we only focus on the time-dependent modulation of GCRs.
Furthermore, for GCRs observed by Voyager 1 \& 2 along their trajectories, 
we only study the lower energy ($\sim 140$ MeV) channel of GCRs.

\begin{figure}
\epsscale{1.} \plotone{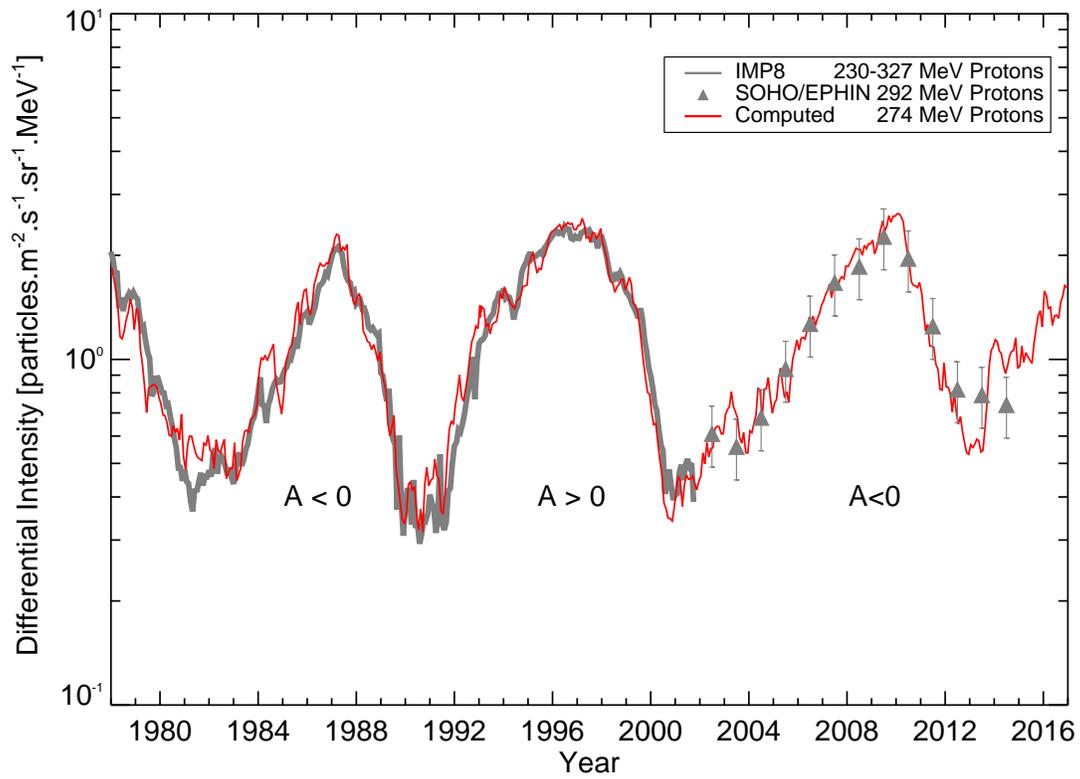}
   \caption{Comparison between the computed monthly 274 MeV proton intensity at 
	Earth (red line) with observations of spacecraft from 1978 to 2016. The gray line 
	means monthly averaged GCR proton intensity, which has been processed to get the 
	GCR background using the despiking algorithm \citep{Qin2012}, from IMP 8 
	observations with energy range in $230-327$ MeV. Gray triangles represent 
	yearly 292 MeV proton flux observed by SOHO/EPHIN \citep{Kuhl2016}.}
   \label{fig:imp81}
 \end{figure}

Figure \ref{fig:imp81} shows the monthly averaged $230-327$ MeV proton 
intensity observed by IMP 8 as a function of time from 1978 to 2001 (gray line) 
and the computed monthly 274 MeV proton intensity (red line) at Earth 
for period 1978-2016. Gray triangles represent yearly 292 MeV proton flux 
observed by SOHO/EPHIN \citep{Kuhl2016}. GCR observations are always contaminated 
by solar cosmic rays \citep[e.g.,][]{ShenAQin2016} which are mainly manifested as 
large spikes in the temporal profile of GCRs. The phase-space thresholding method 
\citep[see,  e.g.,][]{Qin2012} is used to deal with the IMP 8 data for a 
pure GCR background. The profiles of IMP 8 and SOHO/EPHIN data show a clear 
11-year cycle which is anti-correlated with the variations of solar activity. 
Typical peaked and plateau-like time profiles can be seen for $A > 0$ and $A < 0$ 
solar cycles, respectively. During solar maxima, step-like decreases were observed. 
The modulation magnitude in solar maximum 2012-2014 is much smaller than previous 
two solar maxima due to lower solar activity level which can be illustrated by 
the sunspot number from WDC-SILSO (\url{sidc.be/silso/}). 
\citet{Zhao2015} studied the GCR heavy-ion flux in the weak solar 
maximum 2014. Comparing with observations in the last solar maximum 2002, they
found the 
elemental intensities of all heavy nuclei in 2014 to be $\sim$40\% 
higher, which were attributed to the weak modulation associated with 
low solar activity levels in 2014. The computed monthly 
averaged 274 MeV proton intensity at Earth is consistent with spacecraft 
observations. The 11-year cycle and the step-like magnitude during solar 
maximum has been reproduced. The model also gives peaked and plateau-like 
temporal profiles during the $A>0$ and $A<0$ solar minimum, respectively. Small 
peaks on the temporal profile of modeling result are caused by the violent 
changes of input parameters, especially the variation of magnetic turbulence 
magnitude. In the period of 1981-1982, the computed result is higher than the 
observation, which may be caused by the increased perpendicular diffusion 
coefficient as has been shown in the second panel of Figure \ref{fig:lambda}. 
The magnetic turbulence magnitude is relatively low in the first half of 1984, 
which causes a peak in the temporal profile of parallel mean-free path. Thus 
the computed result in 1984 is relatively higher than the observation, and the 
following sharp decrease of the computed result is attributable to large spikes 
exist in IMF and magnetic turbulence. Such phenomenon can also be seen in 1991 
and 2003. The computed result has a sharp increase in 2013, while the observed 
proton intensity keep decreasing until 2014. Despite the relatively larger 
perpendicular mean-free path in 2013, the solar polar field also depicts some 
unusual variation characteristics. From the bottom panel of Figure \ref{fig:field} 
we can see that $P_A=1$ from June 2013 to December 2013, and this can be an 
important factor leading to the sharp increase in GCR intensity. Note that it is 
hard to use our 
model to reproduce the temporal profile of the observed galactic proton 
intensity precisely, because the modeling heliosphere is oversimplified 
(e.g., the Parker HMF and the turbulence model). Nonetheless, the reproduced 
11-year cycle and variation amplitudes over solar maximum can help us understand 
the physical processes of solar modulation. To have a better scenario, we show 
the yearly averaged modeling result in the following work. 

\begin{figure}
\epsscale{1.} \plotone{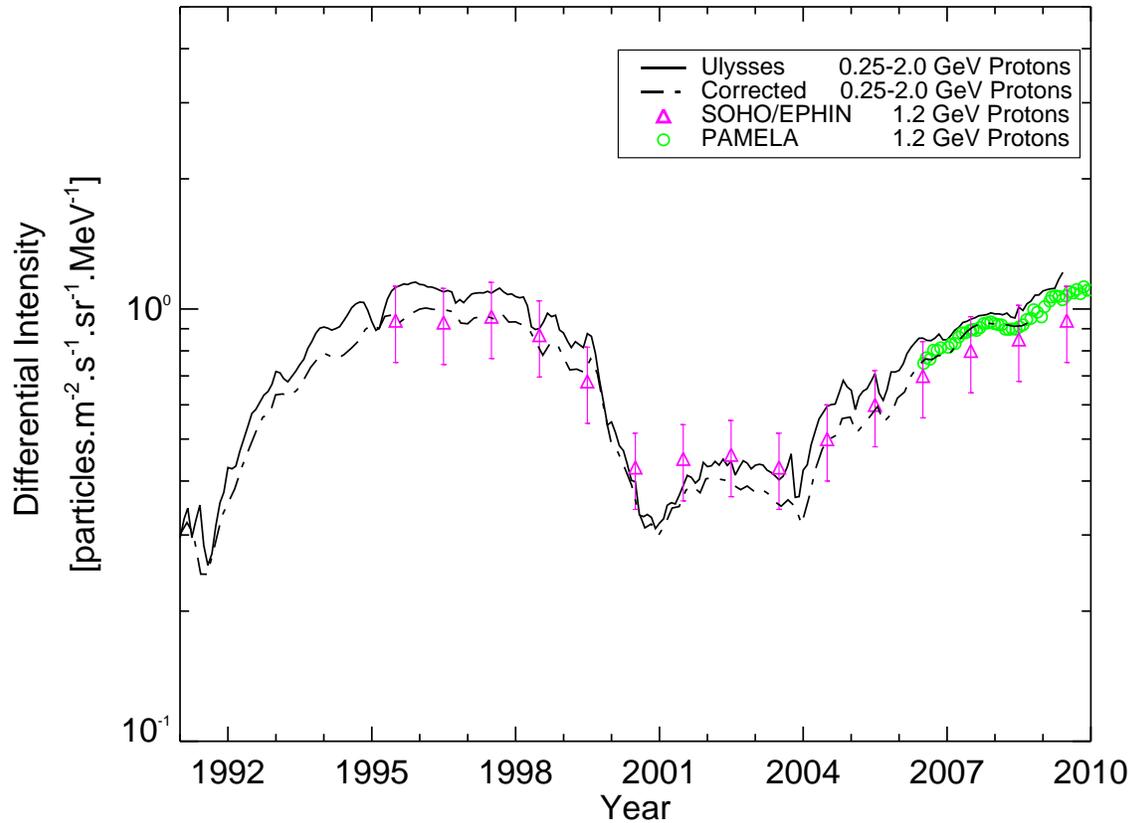}
   \caption{Monthly averaged $0.25-2.0$ GeV proton flux data observed by KET onboard 
	Ulysses is shown as black solid line. The proton intensity data 
	for this channel, which are originally obtained as daily count rates from the 
	Ulysses Final Archive (\url{ufa.esac.esa.int/ufa}), are divided by 
	the corresponding response factor to get the proton flux. The dashed-dotted line means 
	the 1 au equivalent $0.25-2.0$ GeV proton flux, and the relevant count rates 
	are digitized from Figure 5 in \citet{Heber2009}. Note that the raw count rates are 
	corrected by the spatial variations of Ulysses with the spatial gradients of 
	GCR proton intensity \citep{Heber2009} to get the 1 au equivalent count rates. 
	In addition, the 1.2 GeV proton flux observed by SOHO/EPHIN \citep{Kuhl2016} 
	and PAMELA \citep{Adriani2013} are shown as magenta triangles and green 
	circles, respectively.}
   \label{fig:ulysses}
 \end{figure}

Ulysses was launched on October 6, 1990 and orbited around the Sun with the 
latitude varying from $-80^\circ$ to $80^\circ$ and the solar distance ranging from 
$\sim 1$ au to $\sim 5$ au \citep{Heber2009}. The Kiel Electron Telescope (KET) 
on board Ulysses measured electrons in the energy range from $\sim 3$ MeV to above 
300 MeV, and protons and helium nuclei in the energy range from $\sim 5$ MeV/nuc 
to above 2 GeV/nuc \citep{Simpson1992}. The data of KET coincidence channel 
K12, which measures protons with energy in $0.25-2.0$ GeV, has been used to study 
the modulation of GCR outside the solar ecliptic plane in many works 
\citep[see, e.g., ][]{Ndiitwani2005, Vos2016, Boschini2017b}. However, different 
works used different mono-energetic bin to represent this channel, e.g., 1.08 GeV 
\citep{Rastoin1996}, 2.5 GV 
\citep[i.e., 1.73 GeV, ][]{Ndiitwani2005,Heber2009,Manuel2014} and 2.2 GeV 
\citep{Boschini2017b}. As the KET observations are integrated over a large 
energy interval \citep{Simone2011}, perhaps it is better to weight the model 
results of several energy bins with the Ulysses response function and then 
combine them together \citep{Boschini2017b}. \citet{Heber2009} got the 1 au 
equivalent count rates for this channel by correcting the proton intensity 
with the global spatial gradients of GCR protons, and one can get the 1 au 
equivalent GCR proton flux with the corresponding response factor. In addition, 
the precise cosmic-ray spectra measured by the PAMELA instrument can help us 
roughly estimate the effective energy of KET coincidence channel K12. In Figure 
\ref{fig:ulysses}, the black solid line means monthly averaged $0.25-2.0$ GeV 
proton flux observed by Ulysses. Note that the original daily count rates from 
the Ulysses Final Archive (\url{ufa.esac.esa.int/ufa}) are divided by the 
corresponding response factor to get the proton flux. The dashed-dotted line 
represents the 1 au equivalent $0.25-2.0$ GeV proton flux, and the relevant count 
rates are digitized from Figure 5 in \citet{Heber2009}. The 1.2 GeV proton flux 
observed by SOHO/EPHIN \citep{Kuhl2016} and PAMELA \citep{Adriani2013} are shown 
as magenta triangles and green circles, respectively. The 1 au equivalent GCR 
proton flux roughly matches the 1.2 GeV GCR proton observations, and one can
use $1.2$ GeV to represent Ulysses/KET channel K12. Therefore, we will compute 
the 1.2 GeV proton flux along the trajectory of Ulysses to compare with the 
Ulysses K12 measurements. 

\begin{figure}
\epsscale{1.} \plotone{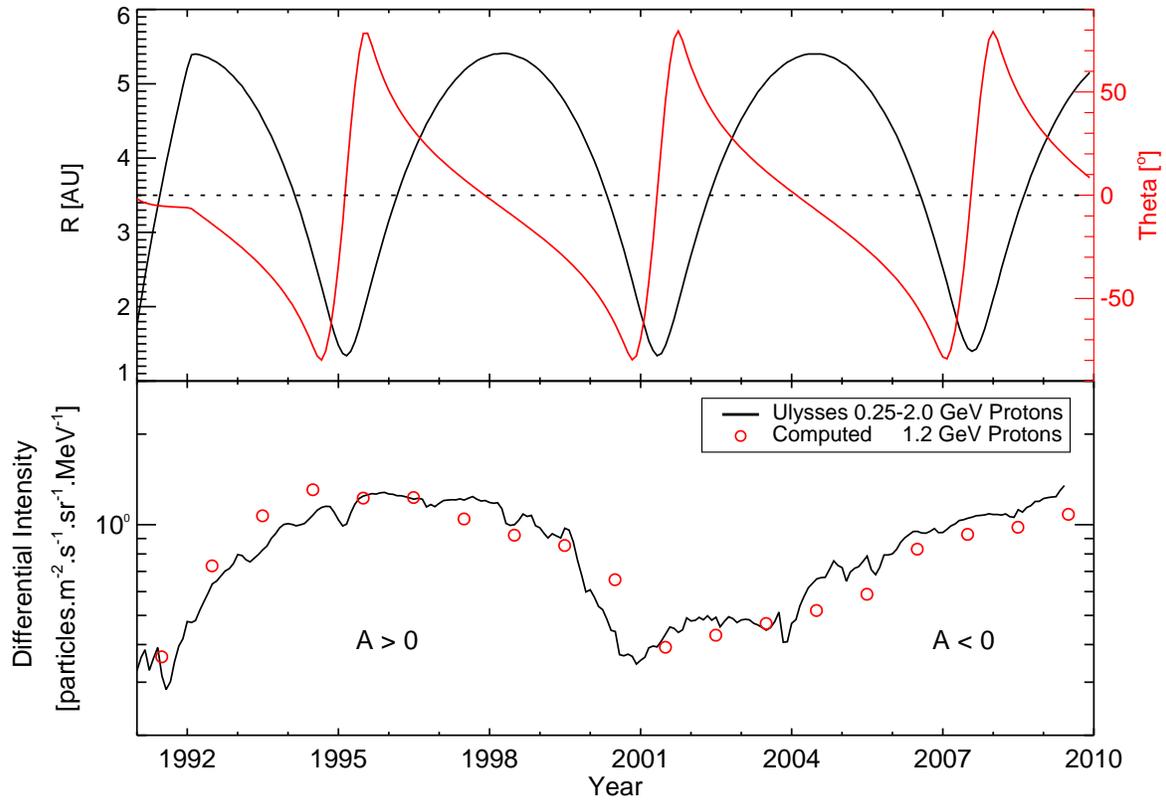}
   \caption{Top panel shows the radial distance (black line) and heliographic 
latitude (red line) of Ulysses. Computed yearly 1.2 GeV proton intensity along 
the Ulysses trajectory science 1991 (red circles) and monthly averaged proton 
intensity observations with energy range in $0.25-2.0$ GeV (black line) from 
1991 to 2009 are illustrated in the bottom panel.}
   \label{fig:ulysses1}
 \end{figure}

Top panel of Figure \ref{fig:ulysses1} illustrates the trajectory of Ulysses. 
Black line is the temporal profile of radial distance varying from 1.4 au to 
5.4 au. Red line means the heliographic latitude of Ulysses as a function of 
time. There are three fast latitude scans when the latitude varying quickly from 
$-80^\circ$ to $80^\circ$ \citep{Heber2009}. Black line in the bottom panel shows 
the intensity of $0.25-2.0$ GeV protons observed by Ulysses/KET, the variations
of which are caused by solar cycles as well as the spatial changes of Ulysses. 
Red circles in the bottom panel are computed yearly averaged 1.2 GeV proton 
intensities along the trajectory of Ulysses. Generally speaking, the numerical 
modeling results are consistent with the observations. The modulation model 
reproduces a plateau-like temporal profile during the $A > 0$ cycle, and the
results during solar maximum are also consistent with observations. However, as we 
consider a mono-energetic bin, the modeling results can not couple with observations 
all the time. In period 2004-2005, modeling results are lower than 
observations 
to some extent. As has been shown in the bottom panel of Figure \ref{fig:input}, 
turbulence in period 2004-2005 is much higher than previous two years, which 
causes smaller diffusion coefficients, and further result 
in a lower computed proton intensity. The modeling results show a sharp increase 
after 2005 due to the obviously decreases in magnetic turbulence, 
magnetic field and tilt angle after 2005. 

\begin{figure}
\epsscale{1.} \plotone{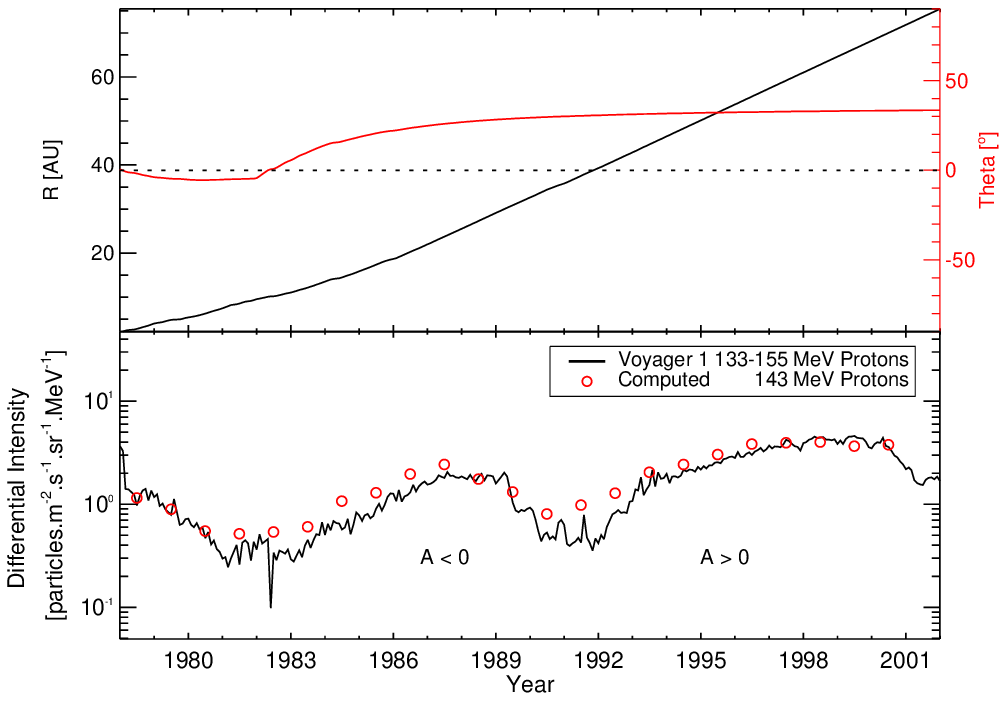}
   \caption{Top panel shows the radial distance (black line) and heliographic 
latitude (red line) of Voyager 1. Computed yearly 143 MeV proton intensity 
along the Voyager 1 trajectory science 1978 (red circles) and monthly averaged proton 
intensity observations with energy range in $133-155$ MeV (black line) from 
1978 to 2000 are illustrated in the bottom panel.}
   \label{fig:voyager1}
 \end{figure}

Figure \ref{fig:voyager1} is another example of numerical results along 
the trajectory of Voyager 1. As the modulation boundary is set at $85$ au, we 
only study the time-dependent modulation from 1978 to 2000. Top panel shows the 
trajectory of Voyager 1, black line means the radial distance, red line is the 
temporal profile of heliographic latitude varying from $-5.5^\circ$ to $33.7^\circ$. 
Bottom panel illustrates the observed $133-155$ MeV and the computed yearly 
averaged 143 MeV proton intensities as a function of time. Note that observation 
data shown in this figure are monthly averaged using daily data from GSFC/SPDF 
OMNIWeb interface (\url{omniweb.gsfc.nasa.gov/}). A peaked temporal profile was 
not observed during the $A < 0$ solar minimum, neither did the observed proton 
intensity show obvious decreases in 1988 as expected, because Voyager 1 experienced 
both corotating merged interaction regions and rarefaction regions during 1988. 
The corotating merged interaction regions and rarefaction regions had equal and 
opposite effects on the variation of cosmic-ray intensities \citep{Burlaga1993}. 
The balance lasted until Voyager 1 encountered a GMIR in 1989, meanwhile, the 
instrument observed a large decrease in proton intensity. These complicated 
structures can not be reproduced by the modified Parker HMF. Step-like decreases 
and increases were observed during solar maxima, and the observed decrease/increase 
amplitudes tend to be smaller while the spacecraft moving toward the modulation 
boundary. Such phenomenon is also reproduced by the model to some extent. 
Modeling results near Earth are larger than observations in periods 1981-1982 
and 1991-1992, and such effects are magnified in the distant heliosphere. Along 
the trajectories of Voyager 1 \& 2, the predicted magnetic turbulence magnitude is 
a little lower than observations in period 1981-1986 
\citep[see Figure 4 in][]{Qin2017}, which might be a reason for the larger 
modeling result during this period. 

\begin{figure}
\epsscale{1.} \plotone{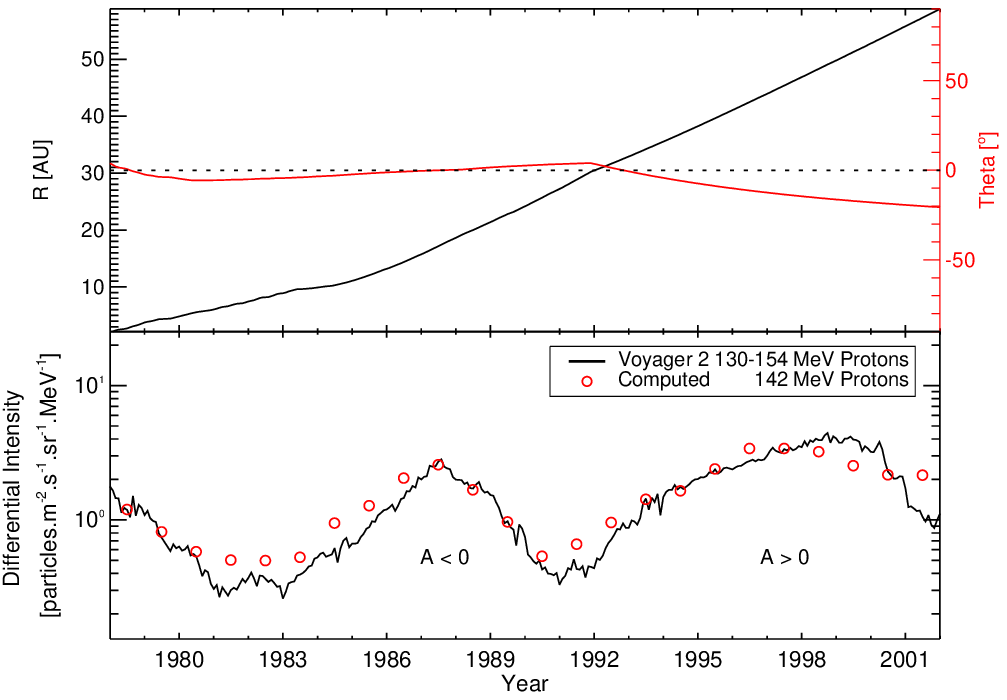}
   \caption{Top panel shows the radial distance (black line) and heliographic 
latitude (red line) of Voyager 2. Computed yearly 142 MeV proton intensity 
along the Voyager 2 trajectory science 1978 (red circles) and monthly averaged proton 
intensity observations with energy range in $130-154$ MeV (black line) from 
1978 to 2001 are illustrated in the bottom panel.}
   \label{fig:voyager2}
 \end{figure}

Figure \ref{fig:voyager2} is similar to Figure \ref{fig:voyager1} except that 
it is for Voyager 2. The heliographic latitude of Voyager 2 varies from 
$\sim -5^\circ$ to $\sim 23^\circ$ and the radial distance ranges from 
$\sim 2$ au to $\sim 65$ au during this period. Note that observation data shown 
in the bottom panel of Figure \ref{fig:voyager2} are monthly averaged 
using daily data from GSFC/SPDF OMNIWeb interface (\url{omniweb.gsfc.nasa.gov/}). 
During the period of 1986-1988 ($A<0$), proton mainly transport along the 
wavy HCS. Voyager 2 moved out near the ecliptic plane while Voyager 1 was above 
the sector zone of HCS \citep{Burlaga2002}, thus proton intensity measured by 
Voyager 2 was larger than that of Voyager 1 \citep{Manuel2011a}. The observed proton 
flux peaked in the period of $1998-1999$ and had a significant decrease which was 
produced by a GMIR in the second quarter of the year 2000 \citep{Burlaga2003}. 
The second step-decrease was observed at the end of the year 2000, produced by 
the arrival of a large GMIR from the Bastille day event at the Earth in 2000 
\citep{Webber2002}. Such step-decreases were also observed by Voyager 1. 
Despite the periods discussed above, the computed results are consistent 
with the spacecraft observations. In the period of 1998-1999, the 
results of our model are lower than observations, and provide a sharp decrease after 
1998, which is about one year later comparing with the modeling results near Earth 
(see Figure \ref{fig:imp81}). The temporal profiles of proton flux observed by 
Voyager 1 and Voyager 2 are nearly the same in the period of 1998-2000. However, the 
profiles of the computed results along the trajectories of Voyager 1 and Voyager 2 
are totally different with the modeling results of Voyager 1 being closer to the 
observations. If a more realistic heliosphere is incorporated in our model, 
the modeling results might fit the observations better. 
\section{Discussion and Conclusions}
In this work, we establish a numerical model to study the modulation of GCRs 
over several solar cycles. As the Parker HMF gives a low magnitude in the polar 
regions at large radial distance, we modify the expression according to 
\citet{Jokipii1989}. New diffusion coefficients model given by \citet{Qin2014} 
is applied. Magnetic turbulence quantities throughout the heliosphere are 
important parameters in this diffusion coefficients model. We only consider the 
variability of magnetic turbulence magnitude $\delta B$ for simplicity, 
which is assumed as a function of spatial location and the tilt angle of HCS 
\citep{Qin2017}. Then we establish a time-dependent 
diffusion coefficients model with some observations at 1 au as the input 
parameters. The effects of HMF modification on particle drifts and diffusion are 
also considered. We assume the Sun's magnetic polarity as random during solar 
maximum with the possibility of $A>0$ decided by the percentage of the north solar 
polar magnetic field being outward and south solar pole magnetic field being 
inward. In addition, a realistic time delayed heliosphere described in 
\citet{Qin2017} is incorporated in our numerical model.

Firstly we study the modulated GCR proton spectra near solar minimum and maximum. 
The computed spectrum near solar maximum turned out to be harder than that of 
solar minimum, which are consistent with observations of IMP8 and BESS, etc. 
However, the proton spectra observed by Voyager 1 and Voyager 2 are hard to be 
reproduced by our model, because they are even softer than the GCR source in our 
model. Thus we focus on the time-dependent modulation of GCRs with single 
individual energy channels instead of the spectra. In addition, we only study 
the low energy channel of GCRs along the trajectories of Voyager 1 \& 2. 

We compute monthly 274 MeV proton intensities at Earth from 1978 to 2016. 
Comparing with the 230-327 MeV proton intensity observations of IMP 8 and 
the observations of SOHO/EPHIN, we believe that our model has reproduced the 
$\sim11$-and $\sim22$-year cycles of solar modulation. The step-like 
magnitude of GCR intensity during solar maximum is also consistent with the 
observations. The 1 au equivalent proton intensity with energy ranges in 
0.25-2.0 GeV (Ulysses K12) from \citet{Heber2009} roughly matches the 1.2 GeV 
proton observations from PAMELA and SOHO/EPHIN measurements. Therefore, 1.2 GeV 
can be used to represent the Ulysses K12 channel. We utilize the numerical model 
to calculate 1.2 GeV proton intensity along the trajectory of Ulysses. Modeling 
results are in close agreement with the observations. Due to the larger magnetic 
turbulence magnitude and smaller radial decay index $S$, the diffusion coefficients 
are much smaller in the heliosphere during the solar maximum and it is harder 
for energetic particles to participate in the inner heliosphere. Thus the numerical 
model gives large decrease during solar maximum. That is also the reason why the 
computed proton intensity is smaller than observations in the period of 
2004-2005. We also study long-term modulation of GCRs in low energy channel 
along the trajectories of Voyager 1 \& 2. The numerical results 
are also consistent with the observations, but the model gets larger results in 
the period of 1981-1986 and 1991-1992, during which perhaps larger turbulence level 
is needed in the distant heliosphere. Besides, from Figure 2 in \citet{Zhao2017}, 
the 2D and slab turbulence energy decrease with the radial distance from 1 to 
75 au. However, because of the pick-up ion driving turbulence, the decay slows 
down after about 10 au.

For the characteristics of magnetic turbulence, we only consider the time 
and spatial dependence of IMF strength and magnetic turbulence magnitude. The 
values of ${\lambda_{slab}}/{\lambda_{2D}}$ and $E_{slab}/E_{total}$ are set as 
constants. Theoretical works \citep[see, e.g., ][]{Oughton2011,Engelbrecht2013a,Zank2017} 
have shown more complicated variation characteristics, and the values of 
${\lambda_{slab}}/{\lambda_{2D}}$ and $E_{slab}/E_{total}$ may changes with solar 
activity. In addition, TRST is a phenomenological model without considering the 
physical processes of turbulence transport in the heliosphere, and the results of 
TRST can not couple with the real condition all the time. Incorporating the 
results of TTMs \citep[e.g., ][]{Zank2017} may help us to study the modulation 
of GCRs. However, it still needs a further study to depict 
turbulence quantities over solar cycles. Modulation effects of heliosheath are 
not considered in this work. Inferring from the work of \citet{Manuel2015}, the 
time-dependent inner heliosheath thickness has significant modulation effects 
on GCR intensities near TS. Thus, incorporating the modulation effects of 
heliosheath and a time-dependent TS position can improve the modeling results 
along the trajectories of Voyager 1 \& 2. Following \citet{Qin2007}, the time 
delayed modified Parker HMF and the solar wind speed are used to establish a 
realistic heliosphere. However, the structure of the real heliosphere is much 
more complicated. Thus the computed results are not consistent with observations 
precisely.

Consequently, the time-dependent diffusion coefficients model based on the work 
of \citet{Qin2014} allow us to study long-term cosmic-ray modulation in the inner 
heliosphere, and all input parameters can be obtained from observations at Earth. 
In the future, we will consider the modulation of GCRs outside the termination 
shock, so we can change the modulation boundary much larger than 85 au and get 
numerical results along trajectories of Voyager 1 \& 2 after 2001. In addition, 
we will study the effects of modulation when particles transport across a sectored 
magnetic field in the out heliosphere \citep{Florinski2011,Florinski2012}. 
Furthermore, we would study the modulation of GCR heavy-ions over solar 
cycles with spacecraft observations and numerical calculations. At last, we would 
work on the study of proton spectra observed by Voyager 1 and Voyager 2 in the 
heliosphere.

\acknowledgments
We are partly supported by grants NNSFC 41374177 and NNSFC 41574172. We used data from 
the Wilcox Solar Observatory 
(\url{wso.stanford.edu}), GSFC/SPDF OMNIWeb interface (\url{omniweb.gsfc.nasa.gov}), 
NASA CDAWeb (\url{cdaweb.sci.gsfc.nasa.gov}) and Ulysses Final Archive 
(\url{ufa.esac.esa.int/ufa}).
The work was carried out at National Supercomputer Center in Tianjin, 
and the calculations were performed on TianHe-1 (A).

\clearpage

\clearpage

\clearpage
\clearpage
\clearpage

\clearpage

\clearpage
\clearpage
\clearpage

\clearpage
\clearpage

\clearpage
\end{document}